\documentstyle[prl,aps,multicol,epsf]{revtex}
\tighten
\narrowtext
\begin{document}
\draft

\title{Semi-classical Theory of Conductance and Noise in Open Chaotic
Cavities}
\author{Ya.~M.~Blanter$^{a}$ and E.~V.~Sukhorukov$^{b,*}$}
\address{
$^a$  D\'epartement de Physique Th\'eorique, Universit\'e de Gen\`eve,
CH-1211 Gen\`eve 4, Switzerland\\
$^b$ Department of Physics and Astronomy, University of Basel,
Klingelbergstrasse 82, CH-4056 Basel, Switzerland}
\date{\today}
\maketitle
\begin{abstract}
  Conductance and shot noise of an open cavity with diffusive boundary
  scattering are calculated within the Boltzmann-Langevin approach. In
  particular, conductance contains a non-universal geometric
  contribution, originating from the presence of open
  contacts. Subsequently, universal expressions for multi-terminal
  conductance and noise valid for all chaotic cavities are obtained
  classically basing on the fact that the distribution function in the
  cavity depends only on energy and using the principle of minimal
  correlations.
\end{abstract}


\begin{multicols}{2}

Transport properties of quantum systems with classically chaotic
dynamics (chaotic cavities) have recently become an object of
intensive investigation. By far the most successful method to describe
them is presently the random matrix theory (RMT), based on assumption
\cite{Baranger,JPB} that the scattering matrix of an open chaotic
cavity is a member of Dyson's circular ensemble of random matrices.
This hypothesis was
quantified to calculate average conductance, weak localization
effects, conductance fluctuations, shot noise, and other quantities
(see \cite{RMP} for review). It has been proven indirectly
\cite{Brouwer} by showing equivalence with Gaussian random matrix
ensembles.

Despite this evident success, a number of questions remains
unaddressed. Thus, by construction all RMT results are {\em
universal}: They only depend on the number of transverse channels in
the leads and the time-reversal symmetry. It is clear, however, that
certain {\em sample-specific} corrections exist. In particular,
average conductance must contain {\em geometric} contributions, since
a cavity connected to open leads is never fully chaotic. Below we show
that these corrections, sensitive to the shape of the cavity, indeed,
exist, and their relative magnitude is of the same order as the
ratio of the cross-section of the leads to the total surface of the
cavity.

Another problem concerns shot noise -- zero-frequency current-current
fluctuations caused by discreteness of electron charge \cite{dJBrev}.
For uncorrelated transfer of classical particles shot noise assumes
the {\em Poissonian value} $S_P = 2e\langle I \rangle$, with $\langle
I \rangle$ being the average current. Fermi statistics induce
correlations, and suppress shot noise below $S_P$. Thus, in metallic
diffusive wires one has $S = 2e\langle I \rangle/3$: The suppression
factor $F = S/S_P$ equals $1/3$. A remarkable feature is that this
$1/3$ suppression was derived in two different ways,
quantum-mechanically \cite{wires1} (based on scattering approach or
equivalent Green's function technique), and semi-classically
\cite{wires2,Jong}, using Boltzmann-Langevin approach
\cite{Kogan}. The equivalence between these two is by no means
evident, and the occurrence of the same result in the two approaches
is even considered as a numerical coincidence
\cite{Landauer}. Furthermore, in symmetric chaotic cavities (equal
numbers of channels in both leads) RMT gives for the
suppression factor \cite{JPB} $F = 1/4$. An alternative, but still
quantum-mechanical derivation of the same result \cite{Nazarov}, as
well as a generalization to the arbitrary two-terminal case
\cite{Nazarov,RMP} and multi-terminal effects \cite{Langen} have been
given. However, it is also desirable to develop a semi-classical
analysis of shot noise in chaotic cavities, based on the
Boltzmann-Langevin approach. Comparison of classical and quantum
results may shed new light onto the physics of shot noise.

Recently, it has been realized \cite{DEK,Louis} that a model of a
circular billiard with diffusive scattering at the boundary (which
models surface disorder on the scale of the wave length) exemplifies
an ``extremely chaotic'' system, with typical relaxation time of the
order of the flight time, and, on the other hand, can be treated
analytically beyond RMT. Ref.\ \cite{DEK} used this example to study
statistical properties of closed cavities. Below we use this model
to study conductance and shot noise semi-classically in cavities
connected to two open leads. In particular, we find corrections to the
conductance which depend on the position of the leads, and reproduce the
$1/4$-suppression of shot noise for the fully chaotic regime. The
crucial observation is that the distribution function of electrons in
the cavity in the leading order is a function only of the energy.

Subsequently, assuming that this is the case for a {\em general}
chaotic cavity, and  using the principle of {\em minimal correlations}
(representing an alternative to the Boltzmann-Langevin method
\cite{Kogan}), supporting the current conservation, we derive the
expressions for the multi-terminal conductance and noise. It is
conceptually important that these expressions obtained
semi-classically reproduce the RMT results available in the
literature.

{\bf Average distribution function}. We consider a circular billiard
of a radius $R$, to which two ideal leads are attached, left (L) at
angles $\theta_0 - \beta/2 < \theta < \theta_0 + \beta/2$, and right
(R) at $-\alpha/2 < \theta < \alpha/2$ (Fig.~1). Eventually, we assume
angular widths of the contacts, $\alpha$ and $\beta$, to be small. In
the semi-classical description, the electron is characterized by
the coordinate $\bbox{r}$,
direction of momentum $\bbox{n}=\bbox{p}/p$,
and energy $E$ (which is
omitted where it can not cause confusion).
Inside the cavity the motion is ballistic, and the average
distribution function $f(\bbox{r},\bbox{n})$ obeys
\begin{equation} \label{Boltzmann0}
\bbox{n} \nabla f(\bbox{r},\bbox{n}) = 0.
\end{equation}
This equation must be supplemented by boundary conditions. At the
surface (denoted $\Omega$) we consider purely diffusive
scattering \cite{Fuchs} for which the distribution function
of the outgoing particles is constant and fixed by flux
conservation,
\begin{displaymath}
f (\bbox{r}, \bbox{n}) = \pi \int_{(\bbox{N}\bbox{n'}) > 0}
\left( \bbox{N} \bbox{n'} \right) f (\bbox{r}, \bbox{n'})
d\bbox{n'}, \ \ \ \left (\bbox{N} \bbox{n} \right) < 0.
\end{displaymath}
\begin{figure}
\narrowtext
{\epsfxsize=4.5cm
\centerline{\epsfbox{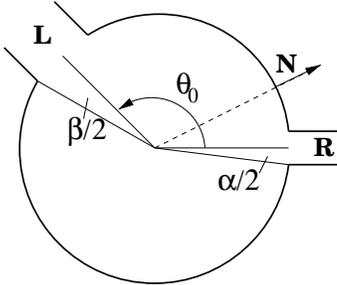}}}
\caption{Geometry of a circular cavity with diffusive boundary
scattering.}
\label{fig1}
\end{figure}
Here $\bbox{r} \in \Omega$, and $\bbox{N}$ is an outward normal to the
surface; we normalized $d\bbox{n}$ so that $\int d{\bbox n} =
1$. Furthermore, electrons coming from the leads are described by
equilibrium distribution functions, $f_{L,R} (E) = \theta(\mu_{L,R} -
E)$, $\mu_L = eV$, $\mu_R = 0$ ($V$ is the applied voltage). We assume
that these incoming particles are emitted uniformly to all
directions. Thus, denoting the cross-section of the right lead
$ \Omega_R = \{ r=R; -\alpha/2 < \theta < \alpha/2 \}$, and
similarly $ \Omega_L$, we complete our system by
\begin{equation}\label{boundary}
f(\bbox{r}, \bbox{n}) = f_{L,R}, \ \ \ \bbox{r} \in
\Omega_{L,R}; \ \ \ \left (\bbox{N} \bbox{n} \right) < 0.
\end{equation}
Due to Eq.\ (\ref{Boltzmann0}), the distribution function inside the
cavity is equal to the distribution function $f(\theta)$ of outgoing
particles, taken at the appropriate point of the surface
$\{ r=R;\theta \}$. For
this function we obtain the integral equation
\begin{equation}
f(\theta)|_{\Omega} =
\frac{1}{4}
\int_{\Omega + \Omega_R + \Omega_L} d\theta'
\left\vert \sin\frac{\theta - \theta'}{2} \right\vert
f(\theta'),
\end{equation}
which has to be solved with the condition
$f(\theta)|_{\Omega_{L,R}} =f_{L,R}$
[see Eq.\ (\ref{boundary})].
This equation applies to arbitrary sizes and positions of the
leads. To progress, we assume the leads to be narrow ($\alpha,
\beta \ll 1$) \cite{narrow}, and
replace integrals of the type $\int_{-\alpha/2}^{\alpha/2} F(\theta)
d\theta$ by $\alpha F(0)$. We then obtain
\begin{eqnarray} \label{distrib}
f(\theta) & = & \frac{\alpha f_R + \beta f_L}{\alpha + \beta} +
\frac{g(0) - g(\theta_0)}{4\pi} \frac{\alpha\beta (\beta -
\alpha)}{(\alpha + \beta)^2} \left( f_L - f_R \right) \nonumber \\
& + & \frac{g(\theta) - g(\theta - \theta_0)}{4\pi}
\frac{\alpha\beta}{\alpha + \beta} \left( f_L - f_R \right),
\end{eqnarray}
with the notation
$$g(\theta) = \sum_{l=1}^{\infty} \frac{\cos l\theta}{l^2} =
\frac{1}{12} \left( 3\theta^2 - 6\pi\theta + 2\pi^2 \right), \ \ \ 0
\le \theta \le 2\pi.$$
The term $(\alpha f_R + \beta f_R)/(\alpha + \beta)$ in
Eq.\ (\ref{distrib}) is uniform and does not depend on $\theta$. Inside
the cavity, it is responsible for a contribution which is position and
momentum independent. It is also the principal term, since all others
are proportional to the first power of $\alpha$ and $\beta$. These
other terms represent {\em geometric corrections}, which are
sensitive to the position of the leads, and, thus, are
sample-specific.

{\bf Conductance}. The average current through the cavity,
$I = e\int_{\Omega_R} d\bbox{r} \sum_{\bbox{p}} \bbox{v}
f(\bbox{r}, \bbox{n}),$
is expressed through the distribution function of outgoing particles,
$f(\theta)$, up to linear order in $\alpha$ and $\beta$,
\begin{equation} \label{current}
I = \frac{eN_R}{2\pi\hbar} \int dE \left\{ \frac{1}{4} \int_{\Omega}
d\theta \left\vert \sin \frac{\theta}{2} \right\vert f(\theta)  - f_R
\right\},
\end{equation}
where we defined the numbers of transverse channels in the left and
the right lead, $N_L = p_FR\beta/\pi\hbar$ and $N_R=
p_FR\alpha/\pi\hbar$. From Eqs.\ (\ref{distrib}) and (\ref{current}) we
obtain the conductance,
$$G = \frac{e^2}{2\pi\hbar} \frac{N_L N_R}{N_L + N_R}
\left[ 1 + \frac{\alpha\beta}{8\pi (\alpha + \beta)} \theta_0
(2\pi - \theta_0) \right].$$
The first term is precisely the result given by RMT, and the second
one represents a non-universal correction, which depends on the
position of the leads $\theta_0$. It is positive and reaches a maximum
for $\theta_0 = \pi$. We stress that this correction is not due to weak
localization effects (which can not be treated classically at all),
but really a geometric correction, originating from the fact that the
escape time from an open cavity is always finite, and the cavity is
never fully chaotic. To the best of our knowledge, these corrections
have never been discussed \cite{Louis1}.

{\bf Shot noise}. Now we give a semi-classical theory of noise
suppression. We only keep the leading terms in $\alpha$ and $\beta$,
though the Boltzmann-Langevin method \cite{Kogan,Jong}, employed
below, can also give geometric corrections. We express the
distribution function as a sum of average (given by Eq.\
(\ref{distrib})) and fluctuating parts. The latter one, $\delta
f(\bbox{x},t)$, obeys the Boltzmann equation,
$$\left( \partial/\partial t + v_F \bbox{n} \nabla \right) \delta
f(\bbox{x},t)  = j(\bbox{x},t),$$
where $\bbox{x} \equiv \{ \bbox{r}, \bbox{n}, E \}$.
The random Langevin source $j$ is zero on average, and its
correlations are
\begin{eqnarray} \label{Lang}
\langle j(\bbox{x},t) j(\bbox{x'},t')
\rangle & = & \nu^{-1} \delta(\bbox{r} - \bbox{r'}) \delta(t-t')
\nonumber \\
& \times & \delta (E-E') J(\bbox{r}, \bbox{n}, \bbox{n'}),
\end{eqnarray}
with $\nu = m/(2\pi\hbar^2)$ being the density of states,
and
%\begin{eqnarray*}
%& & J(\bbox{r},\bbox{n}, \bbox{n''}) = -W_{\bbox{n}, \bbox{n''}}
%\left[ f(1-f'') - f''(1-f)\right] \\
%& & + \delta(\bbox{n} - \bbox{n''}) \int d\bbox{n'}
%W_{\bbox{n},\bbox{n'}} \left[ f(1-f') + f'(1-f) \right],
%\end{eqnarray*}
\begin{eqnarray*}
J(\bbox{r},\bbox{n}, \bbox{n''}) & = &
   \int d\bbox{n'}
   \left[\delta(\bbox{n} - \bbox{n''})-
   \delta(\bbox{n'} - \bbox{n''})\right]
\\
& \times &\left[W_{\bbox{n},\bbox{n'}} f(1-f')+W_{\bbox{n'},\bbox{n}}
f'(1-f)\right],
\end{eqnarray*}
where $W_{\bbox{n},\bbox{n'}}$ is the probability of scattering from
$\bbox{n}$ to $\bbox{n'}$ per unit time at the point $\bbox{r}$. We
introduced the notations $f \equiv f(\bbox{r}, \bbox{n})$, and $f'
\equiv f(\bbox{r}, \bbox{n'})$. Note $\int J d\bbox{n} = \int J
d\bbox{n'} = 0$, due to the conservation of the number of particles.

The probability $W_{\bbox{n}, \bbox{n'}} (\bbox{r})$ can be found from
the following considerations. It is only non-zero for $\bbox{nN} >0$,
$\bbox{n'N} < 0$, and under these conditions does not depend on
$\bbox{n'}$. Thus, $W_{\bbox{n}, \bbox{n'}} = 2\tilde W_{\bbox{n},
\bbox{r}}$, where $\tilde W_{\bbox{n}, \bbox{r}}$ is the probability
of the particle $\bbox{n}$ to be scattered per unit time at the point
$\bbox{r}$. During the time interval $\Delta t$ the particles which
are closer to the surface then $v_F \bbox{nN} \Delta t > 0$ are
scattered with the probability one, and others are not scattered at
all. Taking the limit $\Delta t \to 0$, we obtain
\begin{eqnarray*}
W_{\bbox{n}, \bbox{n'}} = \left\{ \matrix{
v_F (\bbox{nN}) \delta(R - r) , & \bbox{nN} >0, \bbox{n'N} < 0
\cr
0, & {\rm otherwise}. } \right.
\end{eqnarray*}

Now we define the noise spectral power at zero frequency
$\omega =0$ in the usual way,
$S_{ij} = 2\int dt\langle \delta I_i(t) \delta I_j (0)\rangle,$
where $i,j = L,R$, and $\delta I_i (t)$ is the fluctuating current at
lead $i$, associated with the function $\delta f$. After standard
transformations \cite{Jong}, we obtain for $S \equiv S_{RR}$
$$S = 2e^2\nu \int dE \int d\bbox{n} d\bbox{n'}
d\bbox{r} T_R (\bbox{r}, \bbox{n}) T_R (\bbox{r}, \bbox{n'})
J(\bbox{r}, \bbox{n}, \bbox{n'}),$$
where $T_R(\bbox{r},\bbox{n})$ is the probability that the particle at
$(\bbox{r}, \bbox{n})$ will exit through the right lead. It obeys
Eq.\ (\ref{Boltzmann0}) with the boundary condition of diffusive
scattering; in addition, it equals $1$ at $ \Omega_R$ and $0$
at $\Omega_L$. In the leading order $T_R = \alpha/(\alpha +
\beta)$; this order, however, does not contribute to $S$, since $\int
J d\bbox{n} = \int J d\bbox{n'} =0$. The next order is that
for $\bbox{n}$ pointing out to the right (left) contact, $T_R = 1
(0)$. Substituting the average distribution function $f = (\alpha f_R
+ \beta f_R)/(\alpha + \beta)$ into $J$, we obtain
$S=2eGVF$, with the noise suppression factor
\begin{equation} \label{Fano}
F =  \frac{N_LN_R}{(N_L + N_R)^2},
\end{equation}
coinciding with the quantum-mechanical result \cite{Nazarov,RMP}. In
particular, for symmetric system $N_L = N_R$ we recover
$1/4$-suppression, as found previously in Ref.\ \cite{JPB}.

{\bf Multi-terminal conductance}. The rest of the paper, motivated by
Eq.\ (\ref{distrib}), we base on the assumption that in the leading
order the distribution function inside an {\em arbitrary} chaotic
cavity depends only on energy, but not on coordinate or direction of
momentum \cite{assumption}. This assumption should be considered as an
alternative to the assumption that the scattering matrix of the cavity
is taken from Dyson's circular ensemble \cite{Baranger,JPB}. First we
show that this assumption together with the requirement of current
conservation, determines the conductance in the multi-terminal case.

Consider a chaotic cavity of an arbitrary shape, connected through $N$
ideal leads of widths $W_n$ to the reservoirs, described by
equilibrium distribution functions $f_n (E) = f_F(E - eV_n)$, $f_F$
being Fermi distribution function.
It is instructive to define the outgoing current $J_n (E) dE$ through
the $n$th lead in
the energy interval between $E$ and $E+dE$, so that the total current
is $I_n = \int J_n dE$. Denoting the
distribution function in the cavity as $f_C (E)$, we easily obtain
$J_n (E) = e^{-1} G_n \left( f_C - f_n \right)$,
where  $G_n = e^2 N_n/(2\pi\hbar)$ is the Sharvin conductance of the
$n$th contact and $N_n$ is the number of transverse channels. Without
inelastic scattering inside the cavity, the current in each energy
interval must be conserving. From the condition $\sum J_n = 0$ we
immediately find the distribution function $f_C$,
\begin{equation} \label{distrib1}
f_C = \sum_n \alpha_n f_n, \ \ \ \alpha_n \equiv G_n/\sum_n G_n.
\end{equation}
Then, defining the conductance matrix $G_{mn}$ by $I_m = \sum_n G_{mn}
V_n$, we obtain
\begin{equation} \label{conductance}
G_{mn} = \left(\alpha_m - \delta_{mn}\right) G_n.
\end{equation}
This conductance matrix is symmetric, and for the two-terminal case
becomes $G_{LR} = (e^2/2\pi\hbar)(N_LN_R/(N_L + N_R))$, which is the
RMT result.

{\bf Multi-terminal shot noise}.
The principle difficulty of the application of the standard
Boltzmann-Langevin method to the noise of the chaotic cavity is that
this approach \cite{Kogan} uses the collision integral
explicitly. Though we managed above to avoid this difficulty for the
circular cavity with surface scattering, it can hardly be overcome for
an arbitrary ballistic cavity with chaotic dynamics. To resolve this
problem, we formulate the principle of minimal correlations, which is
described below.

The actual origin of shot noise can be viewed semi-classically
as a result of partial occupation of the states by electrons. In the
equilibrium state with finite temperature $T$ the partial
occupation causes fluctuations of the distribution function with the
equal time correlator \cite{Gantsevich}
\begin{equation} \label{equilib}
\langle \delta f(\bbox{x},t) \delta f(\bbox{x'},t) \rangle =
\nu^{-1} \delta (\bbox{x} - \bbox{x'})f(\bbox{x}) \left[ 1 -
f(\bbox{x}) \right].
\end{equation}
The $\delta$-function on the {\em rhs} of this equation means
that the cross correlations are completely suppressed.
We note that in the chaotic cavity the cross correlations should
be also suppressed because of multiple random scattering inside
the cavity. Therefore, we assume now that Eq.\ (\ref{equilib})
is valid for fluctuations of the  non-equilibrium state of the cavity,
where the function $f_C$ plays the role of $f(\bbox{x})$ in Eq.\
(\ref{equilib}). Taking into account that for $t\neq t'$ the
correlator obeys the kinetic equation, $(\partial_t + v_F \bbox{n}
\nabla) \langle \delta f(\bbox{x},t) \delta f(\bbox{x'},t') \rangle =
0$ \cite{Gantsevich}, we obtain the formula
\begin{eqnarray} \label{fluctfun}
& & \langle \delta f(\bbox{x},t) \delta f(\bbox{x'},t') \rangle =
\nu^{-1} \delta \left[ \bbox{r} - \bbox{r'} - v_F \bbox{n} (t-t')
\right] \nonumber \\
& & \times \delta (\bbox{n} - \bbox{n'}) \delta (E-E') f_C (1-f_C),
\end{eqnarray}
which describes strictly ballistic motion and is therefore only valid
at the time scales below the time of flight. On the other hand, after
a time of the order of the dwell time $\tau_d$ the electron becomes
uniformly distributed and leaves the cavity through the $n$th contact
with the probability $\alpha_n$. For times $t\gg\tau_d$ (which are of
interest here) this can be described by an instantaneous fluctuation
of the isotropic distribution $\delta f_C(E,t)$, which is not
contained in Eq.\ (\ref{fluctfun}). The requirement of the
conservation of the number of electrons in the cavity leads to {\em
minimal correlations} between $\delta f_C(E,t)$ and $\delta
f(\bbox{x},t)$ \cite{Lax}. Thus, for the fluctuation of the current at
the contact $n$ we write 
\begin{equation}
\delta I_n  =  \frac{ep_F}{2\pi\hbar^2} \int_{\Omega_n} d\bbox{x}
\ (\bbox{nN}_n) \left[\delta f(\bbox{x},t)+\delta f_C(E,t)\right],
\end{equation}
where $\Omega_n$ and $\bbox{N}_n$ denote the surface of the contact
$n$ and the outward normal to this contact. The requirement that
current is conserved {\em at every instant of time}, $\sum_n \delta I_n
= 0$, eliminates fluctuations $\delta f_C$. After
straightforward calculations with the help of Eq.\ (\ref{fluctfun})
we arrive at the expression
\begin{equation} \label{noisemult}
S_{mn} = -2G_{mn} (T + T_C), \ \ \ T_C = \int dE f_C (1 - f_C).
\end{equation}
Eq.\ (\ref{noisemult}), together with Eq.\ (\ref{conductance}),
constitutes a finite-temperature multi-terminal expression for the
noise power in a chaotic cavity. From Eq.\ (\ref{distrib1}) we obtain
$T_C=(e/2)\sum_{n,m}\alpha_n\alpha_m(V_n-V_m)\coth
\left(e(V_n-V_m)/2T\right)$. At zero temperature this reproduces the
noise suppression factor (\ref{Fano}) and the multi-terminal RMT
results for shot noise \cite{Langen}. At equilibrium, $S_{mn} =
-4TG_{mn}$, in accordance with fluctuation-dissipation theorem.

In conclusion, we presented a semi-classical theory of transport in
chaotic cavities. First, for the model of a circular billiard with
diffusive boundary scattering, we reproduced the RMT results for the
average conductance and shot noise, and, in addition, found geometrical
corrections to the average conductance, which appear due to the fact
that a cavity with open leads is not chaotic any more. Subsequently, we
showed how the RMT results for multi-terminal conductance and noise in
an arbitrary chaotic cavity may be obtained semi-classically
using the principle of minimal correlations.

We thank Y.~Levinson for useful discussions. This work was supported
by the Swiss National Science Foundation. Y.~M.~B.\ thanks The Aspen
Center for Physics, where part of this work was done, for hospitality
and support.

\end{multicols}


\begin{references}
\bibitem{narrow} Note, however, that though width of each lead is much
less than $R$, it is still greater than the wavelength and
supports a large number of transverse channels.
\end{references}
\end{document}